\documentclass[runningheads]{llncs}
\usepackage[T1]{fontenc}
\usepackage{graphicx}
\usepackage{booktabs}
\usepackage[misc]{ifsym}
\usepackage{tcolorbox} 
\usepackage{adjustbox} 
\usepackage{mwe}
\usepackage{booktabs}
\usepackage{amsmath}
\usepackage{adjustbox} 
\usepackage{enumitem}
\usepackage{xcolor}
\usepackage{tcolorbox} 
\usepackage{multirow}
\usepackage{multicol}
\usepackage{colortbl}
\usepackage{CJKutf8}

\usepackage{mwe}
\begin{document}
\title{FinCPRG: A Bidirectional Generation Pipeline for Hierarchical Queries and Rich Relevance in Financial Chinese Passage Retrieval}

\titlerunning{FinCPRG: A Bidirectional Generation Pipeline in Passage Retrieval}


\author{
Xuan Xu\inst{1} \and
Beilin Chu\inst{1} \and
Qinhong Lin\inst{1} \and
Yixiao Zhong\inst{1} \and
Fufang Wen\inst{2} \and
Jiaqi Liu\inst{2} \and
Binjie Fei\inst{2} \and
Yu Li\inst{2}\Letter \and
Zhongliang Yang\inst{1}\Letter  \and
Linna Zhou\inst{1}\Letter \def\thefootnote{*}\footnote{Equal Corresponding Authors: YL, ZY and LZ.}
}

\tocauthor{Xuan Xu, Beilin Chu, Qinhong Lin, Yixiao Zhong, Fufang Wen, Jiaqi Liu, Binjie Fei, Yu Li, Zhongliang Yang, Linna Zhou}
\toctitle{FinCPRG: A Bidirectional Generation Pipeline for Hierarchical Queries and Rich Relevance in Financial Chinese Passage Retrieval}


\authorrunning{Xuan Xu et al.}
\institute{
Beijing University of Posts and Telecommunications, Beijing 102206, China \email{\{sh22xuxuan,beilin.chu,greenred99,050922zyx,yangzl,zhoulinna\}@bupt.edu.cn}
\and
Beijing Value Simplex Technology Co. Ltd, Beijing 100026, China \email{\{wenfufang,liujiaqi,feibj,liyu\}@entropyreduce.com}
}

\maketitle              


\begin{abstract}
In recent years, large language models (LLMs) have demonstrated significant potential in constructing passage retrieval datasets. However, existing methods still face limitations in expressing cross-doc query needs and controlling annotation quality. To address these issues, this paper proposes a bidirectional generation pipeline, which aims to generate 3-level hierarchical queries for both intra-doc and cross-doc scenarios and mine additional relevance labels on top of direct mapping annotation. The pipeline introduces two query generation methods: bottom-up from single-doc text and top-down from multi-doc titles. The bottom-up method uses LLMs to disassemble and generate structured queries at both sentence-level and passage-level simultaneously from intra-doc passages. The top-down approach incorporates three key financial elements—industry, topic, and time—to divide report titles into clusters and prompts LLMs to generate topic-level queries from each cluster. For relevance annotation, our pipeline not only relies on direct mapping annotation from the generation relationship but also implements an indirect positives mining method to enrich the relevant query-passage pairs. Using this pipeline, we constructed a Financial Passage Retrieval Generated dataset (FinCPRG) from almost 1.3k Chinese financial research reports, which includes hierarchical queries and rich relevance labels. Through evaluations of mined relevance labels, benchmarking and training experiments, we assessed the quality of FinCPRG and validated its effectiveness as a passage retrieval dataset for both training and benchmarking.\footnote{https://github.com/valuesimplex/FinCPRG}

\end{abstract}

\section{Introduction}

In the domain of dense retrieval (DR), passage retrieval datasets are typically composed of three fundamental elements: queries, passages, and relevance labels.
Due to the high cost of collecting queries and relevance labels \cite{wang2024csprd}, as well as challenges in ensuring quality and diversity, traditional methods heavily rely on internal search business data accumulation and crowdsourcing \cite{xie2023t2ranking,qiu2022dureader_retrieval}, as well as dataset collection \cite{thakur2021beir,muennighoff-etal-2023-mteb,tang2025finmtebfinancemassivetext}, leading to slow progress in low-resource and highly specialized scenarios.
In recent years, LLMs have revolutionized traditional data engineering practices, such as (semi-)automated dataset construction, due to their strong generalization capabilities in natural language processing tasks \cite{brown2020language,schickItNotJust2021}. A surge of work has emerged around LLM-based synthetic data generation, covering pretraining corpora, instructions, question-answering, and other data types. Correspondingly, many studies \cite{weller-etal-2024-generative} have attempted to synthesize and augment passage retrieval datasets using LLMs. These methods can be classified according to focus on synthesizing or augmenting one of the key elements, specifically: a) Query or Document synthesis and augmentation \cite{jeronymo2023inpars,gao-etal-2023-precise,wang2023query2doc,rahmani2024synthetic}; b) Automatic Labeling such as Positives or negatives relevance label mining \cite{wang2021gpl,rahmani2024synthetic,chuxin2024embedding}. These approaches have demonstrated the potential of synthetic data in terms of effectiveness.
However, there are two core issues in using LLMs to synthesize high-quality retrieval datasets.
First, due to the limited context window of LLMs, generated queries are often confined to individual documents, lacking global information across the document set, and failing to synthesize complex queries spanning multiple paragraphs (e.g., issues exposed in datasets like \cite{wang2023query2doc,chen-etal-2024-fintextqa,chuxin2024embedding}).
Second, previous designs of synthesis pipelines are relatively simplistic, failing to comprehensively leverage traditional methods and various neural models, resulting in insufficient quality and diversity control \cite{wang2023improving,zhang-etal-2023-contrastive-learning}. Especially, these models are not tailored to any specific domain, which may result in the overlooking of key semantic elements pertinent to the domain. Consequently, this can lead to a relatively high rate of false positives or false negatives \cite{chen2024airbench} in relevance labels. For instance, in the financial sector, elements such as industry, company entities, and time play crucial roles.


To address these challenges, we propose a bidirectional generation pipeline for automated passage retrieval dataset construction. This pipeline combines bottom-up (for intra-doc queries) and top-down (for cross-doc queries) approaches to generate hierarchical queries while employing an indirect positives mining method to balance efficiency and coverage of relevance label mining.

The core of the bottom-up approach for intra-doc queries is generating sentence-level and passage-level queries simultaneously given a specific passage of a document, which are then completed and hierarchically organized. 
The preprocess process involves segmenting document text into paragraphs and sentences. Subsequently, low-quality passages are filtered using a BERT-based quality scorer and regex rules. Then we use LLMs to disassemble and generate structured queries at both levels simultaneously. Subsequently, extract company entity names from metadata to resolve issues of ambiguous references in the generated query. Finally, we organize queries from different paragraphs within the same document into larger hierarchical query sets for subsequent indirect relevance mining.

The top-down approach for cross-doc queries is inspired by human report reading processes: people typically approach document collections with specific interests and intentions within certain industries and topics, scanning index-type text (usually chapter titles or paragraph topic sentences) to locate relevant passages for detailed reading. Building on this insight, we leverage industry classification models and topic modeling techniques to divide clusters for report titles, incorporating three core semantic dimensions: industry, topic, and temporal factors. Subsequently, we utilize LLM to generate the intention to consult the documents and then decompose it into fine-grained query sets, guided by a list of representative titles within each cluster.

What's more, we propose a comprehensive automatic relevance annotation strategy combining direct mapping annotation with indirect positives mining. The direct mapping annotation, similar to previous work, is derived directly from the generation relationships of LLMs, where queries are mapped to their source passages.
 However, relying solely on direct associations overlooks the relevance between the generated queries and other adjacent passages. To address this, we indirectly determine the relevance between queries and passages through localized traversal of query pairs using a reranker, which identifies a large number of additional relevance labels, alleviating the false negative issue in synthetic datasets.
We sampled approximately 1,300 Chinese financial research reports across 19 categories, including company reports, industry reports, and fund reports. Using our pipeline, we synthesized queries at three granularity levels, ultimately constructing FinCPRG, a financial Chinese passage retrieval generated dataset comprising five subsets (sentence, sentence-mined, passage, passage-mined, and topic).

We conducted various experiments on FinCPRG: first analyzing the quality of relevance labels through interval sampling inspection, then evaluating common Chinese open-source retrieval models using FinCPRG as a test set. The evaluation results showed high consistency with two true financial retrieval benchmarks, validating the synthetic dataset's utility as an evaluation set. Additionally, we explored the framework's potential for generating effective training data for low-resource domains. We use FinCPRG as a training set and test on a third-party financial retrieval benchmark. We significantly observe an improvement in its financial domain retrieval capabilities.

Specifically, our contributions are:
\begin{enumerate}
    \item We propose a bidirectional generation pipeline that combines both bottom-up and top-down approaches for generating intra-doc and cross-doc queries. Additionally, we introduce an automated positives mining method to construct a rich annotated dataset while ensuring a balance between efficiency and coverage.
    \item We construct the FinCPRG dataset using our proposed pipeline—a comprehensive Chinese financial passage retrieval dataset. It is derived from approximately 1,300 financial research reports sampled from a collection of 17 types of reports. The dataset features hierarchical queries along with rich relevance labels.
    \item We evaluate the FinCPRG dataset through multiple types of experiments, assessing relevance labels' quality and validating its effectiveness as both an evaluation benchmark and a training dataset.
\end{enumerate}

\begin{figure}[t]
    \centering  
    \includegraphics[width=1\textwidth, 
        trim={1cm 5.6cm 1cm 1cm},  
        clip    
    ]{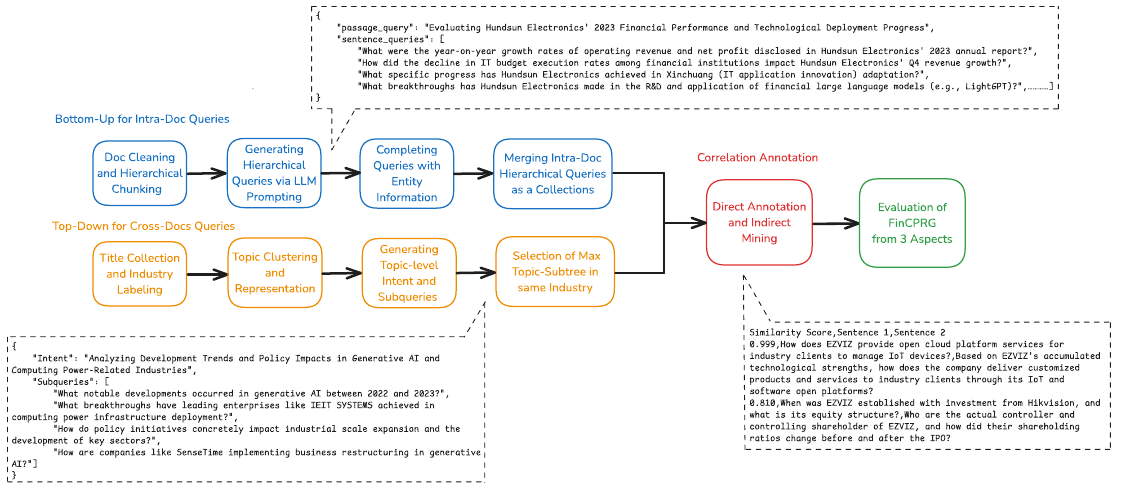}
    \caption{Overview of the FinCPRG pipeline along with intermediate examples.}
    \label{fig1}
\end{figure}

\section{Related Work}

\subsection{Fully Synthetic Datasets for DR}
"Fully synthetic" refers to synthesizing retrieval datasets from raw corpora alone, or even without any corpora. It typically involves the complete pipeline of query synthesis, relevance label mining, quality control, and other processes. E5~\cite{wang2023improving} uses a two-step prompt template to generate synthetic data, first prompting GPT-4 to brainstorm retrieval tasks, then generating (query, positive, hard negative) triplets for each task. In contrast, AIR-Bench~\cite{chen2024airbench} generates characters who would find the document useful and scenarios where they might use it, then creates queries based on specific characters and scenarios, using embedding models, multiple rerankers, and LLMs for quality control. Chuxin-Embedding~\cite{chuxin2024embedding} generates queries similarly with E5 after a choice of the roles, scenarios, and categories and refines query quality using the LLMs, hard negative mining, and reranking techniques. Additionally, Rahmani et al.~\cite{rahmani2024synthetic} validate the reliability of fully synthetic datasets, demonstrating their effectiveness for retrieval evaluation. However, these synthetic datasets have the disadvantages of the absence of cross-document queries and the challenging trade-off between cost and quality.

\subsection{Partly Synthetic Datasets (Data Augmentation) for DR}
Partly synthetic typically refers to original data that contains some elements of retrieval datasets, such as queries and partial relevance labels. It uses LLMs for data augmentation to improve retrieval systems, including query rewriting, label mining/cleaning, document enhancement, and so forth. Works like Inpars-v2~\cite{jeronymo2023inpars} and Query2doc~\cite{wang2023query2doc} leverage LLMs for query and document synthesis, while Precise Zero-Shot Dense Retrieval~\cite{gao-etal-2023-precise} proposes to pivot through Hypothetical Document Embeddings (HyDE). Another critical aspect of synthetic techniques is the generation of relevance labels. Generative Pseudo Labeling (GPL)~\cite{wang2021gpl} introduces an unsupervised method for pseudo-relevance label generation. These approaches reduce reliance on costly human annotations and scale retrieval systems efficiently. However, these methods are fragmented and scattered, requiring integration and adaptation to specific domains.

\section{Methodology}

\subsection{Bottom-Up for Intra-Doc Queries}
The bottom-up approach operates on the passages which are chunked from the doc, leveraging the language understanding and generation capabilities of LLMs to synthesize sentence-level and passage-level queries simultaneously, whose semantic connotations are precisely aligned with the original text chunks. By carefully prompting LLMs to generate queries of two different granularities, comprehensive coverage of all possible queries related to the passage is ensured.

\subsubsection{Document Cleaning and Hierarchical Chunking}
Our initial step involves passage-level segmentation (500 character length) of research reports, employing rule-based matching to eliminate passages containing extensive tables and privacy information such as organization and personal introduction. Given that our original research report data was obtained through PDF recognition, we still encountered various noise elements, including URLs, table/figure captions, and inadvertently recognized text from images. To address this, we annotated a set of low-quality samples and fine-tuned BERT to automatically detect and remove low-quality passages. After cleaning (removing approximately 15\% of the content), we performed secondary chunking (100 length) on the remaining passages, segmenting each passage chunk into sentence-level sub-chunks. This hierarchical chunking structure facilitates the subsequent generation of hierarchical queries.

\subsubsection{Generating Hierarchical Queries via LLMs Prompting}
We designed a complex LLM prompt to generate hierarchical inner-doc queries. Specifically, given an input document chunk, the model generates queries at two distinct granularity levels: passage-level queries that capture broader thematic content and sentence-level queries that focus on fine-grained information. 
It takes full advantage of LLM's instruction-following capabilities to perform more challenging tasks (instead of generating a list of queries, we're more likely to generate a query tree), thus reducing LLM's inference costs. What's more, creating queries at different levels facilitates indirect relevance mining among queries in the same level, as explained in Chapter 4.
The specific prompt design for query generation is as follows:
\begin{tcolorbox}[colback=gray!20, colframe=gray!100, sharp corners, leftrule={3pt}, rightrule={0pt}, toprule={0pt}, bottomrule={0pt}, left={2pt}, right={2pt}, top={3pt}, bottom={3pt}]
{\small 
\textbf{Prompt:} Given a passage from a financial report (provided as a list of sentences), generate hierarchical queries including both passage-level and sentence-level queries. Follow these requirements strictly and return results in JSON format.

\textbf{Input:} \texttt{["Sentence 1.", "Sentence 2.", ..., "Sentence N."]}

\textbf{Requirements:}
1. Ignore disclaimers, copyright notices, or sensitive information
2. Include passage-specific information (company names, events, data)
3. Use empty string ("") for unclear sentences
4. Return in specified JSON format

\textbf{Output Format:} \texttt{\{"passage\_query": "query 0", "sentence\_queries": ["Query 1", "Query 2", ..., "Query N"]\}}
}
\end{tcolorbox}

\subsubsection{Query Completion and Merging}
Some generated queries use fuzzy references as their subject. For example, 'How is the company's financial situation this year?' is an invalid query that requires entity completion. Therefore, we have introduced a query rewriting mechanism to address the issue of incomplete entity names. We employ regular expressions to extract company names from report document titles and use these to replace ambiguous references in the synthesized queries (e.g., replacing generic terms like "company" with specific names like "XX Technology"). After query completion, we collect and organize queries from different passages within the same document to create merged intra-doc hierarchical query collections.

\subsection{Top-Down for Cross-Doc Queries}
The top-down approach for cross-doc queries draws inspiration from human reading behavior: readers typically approach document collections with specific interests and intentions within particular industries or topics, scanning index-type text (such as chapter titles or topic sentences) to locate relevant passages for detailed reading. Following this insight, we leverage industry classification models and topic modeling techniques to divide clusters for report titles, incorporating three key semantic elements: industry, topic, and temporal factors. We then prompt LLMs to generate intentions as topic-level queries based on representative documents and decompose these intentions into fine-grained subqueries simultaneously that will be used in follow-up relevance mining.

\subsubsection{Title Collection and Industry Labeling}

During the data preprocessing phase, we extracted report titles from the metadata of sampled research reports and filtered out titles with fewer than five Chinese characters (e.g., "Daily Morning Report" or "Morning Meeting Digest"), as such short titles typically lack substantive thematic information. Then we performed deduplication to retain only one instance. Subsequently, we employed FinBERT2-IC, an industry classifier fine-tuned on FinBERT2 \cite{xu2025finbert2} to annotate each title with its corresponding industry label. This classifier adheres to the CITIC Securities primary industry classification standard, encompassing 28 distinct industry categories.

\subsubsection{Topic Clustering and Representation}
 This part builds on the BERTopic framework, which is a popular topic modeling tool. Rather than using a general embedding model to extract vectors, which would overlook subtle yet critical differences leading to high similarity— for example, deeming the two titles “Company A is growing rapidly” and “Company B is growing rapidly” as highly similar when, in reality, Company A and Company B belong to different industries—we used a fine-tuned industry classification model, FinBERT2-IC, to encode report titles into embeddings that capture both semantic and industry-specific information.
 
 Furthermore, to incorporate temporal information from report metadata into the clustering representations, we implement a periodic temporal encoding scheme inspired by Time2Vec \cite{kazemi2019time2vec}. This approach maps temporal information into a continuous vector space while preserving the periodic nature of temporal patterns, specifically by encoding the temporal displacement between each document's timestamp and a reference date into a 6-dimensional vector, which is concatenated into the previous embeddings. 
 
\begin{figure}[t]
\includegraphics[width=\textwidth,trim={2cm 5cm 1cm 1cm},clip]{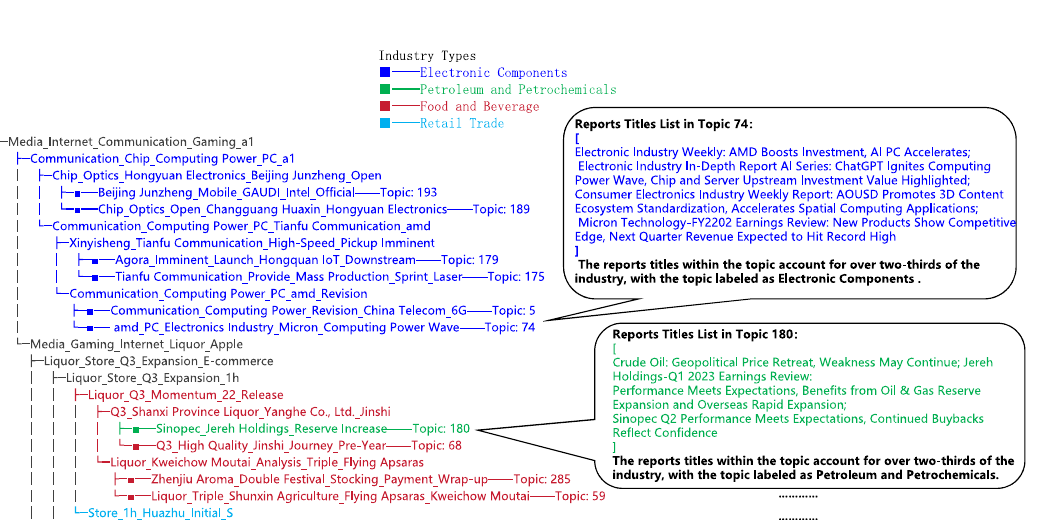}
\caption{The demonstration of the topic tree from the Topic Clustering and Representation phase, highlighting the hierarchical organization of topics. Each topic comprises a set of research report titles. Additionally, two annotation examples of the dominant industry within a specific topic are provided during the phase of selection of the maximum topic subtree within the same industry.} \label{fig2}
\end{figure}

 Based on these embeddings, we apply the HDBSCAN (Hierarchical Density-Based Spatial Clustering of Applications with Noise) algorithm for unsupervised clustering to construct hierarchical clusters. Consistent with BERTopic, we employ c-TF-IDF (Class-Based Term Frequency-Inverse Document Frequency) to identify statistically significant keywords within each cluster and take the keywords list as the topic description. The demonstration of the topic tree from the Topic Clustering and Representation phase is shown in Figure~\ref{fig2}.

\subsubsection{Generating Topic-level Intent and Subqueries from Representative Documents}
We then prompt LLMs to generate intentions to read these title clusters as topic-level queries and decompose these intentions into fine-grained subqueries. Specifically, topic keywords and titles most representative of the topic are passed as input to generate a core intent (summarizing the query purpose) and associated subquery sets simultaneously. These subqueries form a logical chain that systematically deconstructs the users' intent, guiding them to construct meaningful query paths for retrieving the desired passages. 
The translated prompt template is structured as follows:
\begin{tcolorbox}[colback=gray!20, colframe=gray!100, sharp corners, leftrule={3pt}, rightrule={0pt}, toprule={0pt}, bottomrule={0pt}, left={2pt}, right={2pt}, top={3pt}, bottom={3pt}]
{\small 
I have a topic described by the following keywords: \textbf{[KEYWORDS]}
For this topic, the following documents represent a small but representative subset of all relevant documents: \textbf{[DOCUMENTS]}
Please generate a formatted dictionary list representing query intents and sub-query sets related to this topic. 

\textbf{Output Format:}
[\{
    "intent": "{brief description of the core purpose for querying this topic}",
    "subqueries": [
        "{subquery 1: first question to address the intent}",
        "{subquery 2: second question to address the intent}",
        ...]
\}]
}
\end{tcolorbox}

\subsubsection{Selection of Maximum Topic Subtree within Same Industry}
Although we generated topic queries with the most fine-grained cluster partitions previously, there is a topic over-segmentation problem, which is not conducive to the subsequent sufficient indirect mining. So we identify the largest subtree that shares the same industry type as a new cluster unit.
Specifically, we analyze the industry distribution of document titles within each topic. A topic is labeled with a dominant industry if more than two-thirds of its document titles belong to that industry; otherwise, it is labeled as 'none'. If even the smallest topic does not share the same industry as the title, we default to selecting the smallest topic as its maximum topic subtree. 
We select the largest corresponding topic subtree sharing the same industry for each title to prepare for subsequent positives mining.

\section{Correlation Annotation}


\subsection{Positive Annotation Strategy}
We first distinguish three types of relationships between queries and passages:
\begin{enumerate}
    \item Subset Query (sentence-level queries, Q<D): The document contains all information required by the query, along with additional redundant information
    \item Equivalent Query (passage-level queries, Q=D): The document and query contain approximately equivalent information
    \item Superset Query (topic-level queries, Q>D): Multiple documents are needed to fully cover the information required by the query
\end{enumerate}
Our annotation strategy combines direct mapping annotation with the indirect positives mining method. The direct mapping annotation is based on LLM-based hierarchical query generation: passage-level queries naturally correspond to equivalent relationships (Q=D), while sentence-level queries correspond to subset relationships (Q<D). In addition to direct mapping annotation, there are still two broad cases of possible positives that exist. Firstly, both passage-level and sentence-level queries may still have relevant passages in adjacent sections. Secondly, since prompts for generating topic-level queries only incorporate information from titles, the generated queries cannot be directly mapped to specific passages. Therefore, we introduce the indirct positives mining method, whose core is to indirectly determine the relevance between queries and passages through localized traversal of query pairs using a reranker model. 
\subsection{Positives Mining via Localized Traversal between Queries}
 Our indirect positive mining method adopts a localized traversal strategy using a reranker to mine equivalent query pairs. Different from the embedding model, the reranker can get a relevance score by taking two pieces of text as input. To balance efficiency and coverage of the mining process, we set different traversal spaces for queries of different granularities. Then we identify equivalent pairs through threshold filtering, which is set empirically at 0.99. This space constraint also helps mitigate false positive labels caused by semantically similar queries that differ in key semantic elements. The three levels of traversal space are defined as follows:
\begin{itemize}
\item For each sentence-level query, the reranker evaluates its similarity with sentence-level queries from other passages within the same document.
 \item For each passage-level query, the reranker evaluates its similarity with other passage queries under the same maximum topic subtree within the same industry.
\item For each topic intent, the reranker evaluates the similarity between subqueries decomposed from the topic intent and the passage-level queries within the same topic hierarchy.

\end{itemize}

 Compared to previous asymmetric query-document mining methods, our symmetric query-pair mining approach offers several advantages:
\begin{enumerate}
    \item While document texts are cleaned, their quality remains uncertain, whereas queries are more concise with less noise.
    \item Compared to query+document pairs, query pairs are shorter, facilitating more efficient model inference and easier manual verification.
    \item The reranker will perform full-attention over the input pair, which is more accurate than embedding model (i.e., bi-encoder) but more time-consuming. As a result, a localized traversal strategy addresses the time complexity constraint.
\end{enumerate}

\section{Implementation Setup and Results}

\begin{figure}[t]
    \centering  
    \includegraphics[width=\textwidth, 
        trim={1cm 7cm 1cm 1cm},  
        clip     
    ]{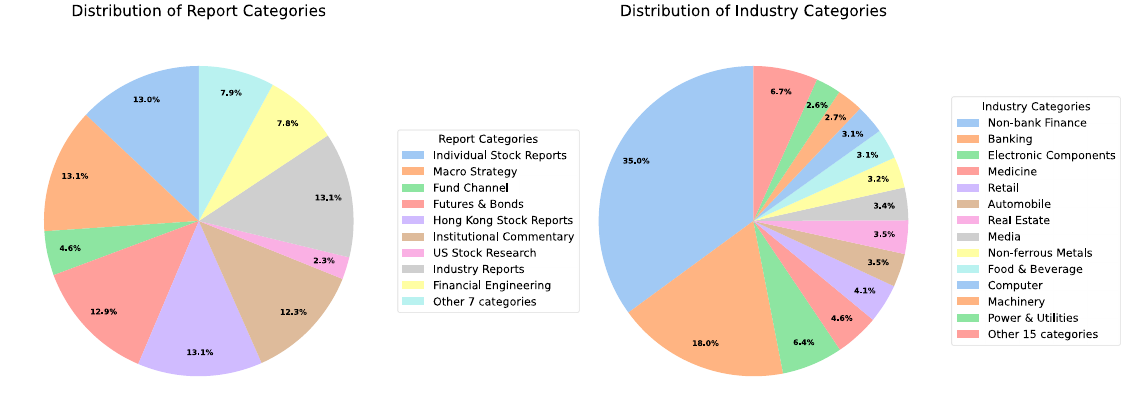}
    \caption{ Pie charts illustrating the distribution of report types (left) and industry categories (right) in the seed research reports dataset.}
    \label{fig3}
\end{figure}

\subsection{Raw Dataset}
To ensure the diversity of the synthetic dataset, we sampled from a research reports collection spanning from 2022 to January 2024. These reports cover subcategories of more than twenty types, including individual stock analysis, industry analysis, futures, and bonds. For each report type, we conducted independent random sampling, with sampling rules set to extract a minimum of 10 samples per category (or all samples if less than 10) and a maximum of 200 samples. Finally, we obtained a dataset containing 1,317 research reports. The categories and industry distribution of the raw dataset are shown in the figure~\ref{fig3}.

\subsection{Implementation Details}
Any steps that involve using LLM are implemented by calling the gpt-4o-2024-11-20 API. Industry classification tasks were performed using FinBERT2-IC\footnote{\url{https://github.com/valuesimplex/FinBERT2}}. The hierarchical clustering and topic representation in our pipeline were implemented based on the BERTopic\footnote{\url{https://github.com/MaartenGr/BERTopic}}, and the generation of intentions and subqueries also utilized the LLM-based custom topic representation module in the BERTopic library. For indirect positives mining, we employed the BGE-reranker-v2-m3\footnote{\url{https://huggingface.co/BAAI/bge-reranker-v2-m3}} from the BGE (BAAI General Embedding) series. 

\begin{table}[t]
    \centering
    \footnotesize  
    \caption{Statistic of FinCPRG, FinCPRG-all is the collection of other five subsets.}
    \label{tab:tsv_stats}
\begin{adjustbox}{width=0.8\textwidth, center}
    \begin{tabular}{lcccr}
        \toprule
        \textbf{Subset Name} & \textbf{Avg. Query} & \textbf{Avg. Doc.} & \textbf{Avg. Rel. Docs} & \textbf{Count of} \\
        & \textbf{Length} & \textbf{Length} & \textbf{per Query} & \textbf{Pairs} \\
        \midrule
        FinCPRG-sentence & 37.18 & 448.70 & 1.00 & 45,457 \\        FinCPRG-sentence-mined & 39.95 & 457.73 & 2.65 & 19,464 \\

        FinCPRG-passage & 45.43 & 564.47 & 1.00 & 10,624 \\
        FinCPRG-passage-mined & 46.14 & 635.25 & 9.06 & 18,216 \\
        FinCPRG-topic & 21.84 & 579.39 & 2.95 & 1,113 \\
        FinCPRG-all & 38.63 & 564.47 & 1.71 & 94,874 \\
        \bottomrule
    \end{tabular}
\end{adjustbox}
\end{table}

\subsection{Final Dataset Statistics}
By integrating the aforementioned synthesis and annotation methods, we obtained five types of query-passage relevance labels. Among these, sentence-level and passage-level were direct mapping annotated, while sentence-level-mined, passage-level-mined, and topic-level were derived through indirect positives mining. Along with the passages from all of our research reports as the corpus, we ultimately constructed five paragraph retrieval datasets. Statistics of each subset in FinCPRG are shown in Table~\ref{tab:tsv_stats}.

\section{Experiments and Evaluation}

\subsection{Evaluation of Mined Relevance Labels}
\subsubsection{Methods}

Although we used a localized traversal method to evaluate the relevance of each query pair in the search space, our arbitrary threshold-setting approach for determining relevance and the label quality require further analysis. Therefore, we first analyzed the distribution of similarity scores from the reranker to gain insights into the model's scoring preferences. Subsequently, we designed an evaluation framework based on both LLM and human assessments to further measure the quality of similarity scores obtained from the reranker. While cost considerations prevented us from employing LLM and human methods during the mining process, we sampled a subset of data for evaluation purposes to provide valuable insights into the reasonableness of our threshold settings and the quality of our mined labels.




In the evaluation of query pairs scored by the reranker with different similarity, we randomly sampled 50 pairs from 8 high-similarity intervals (0.99–1.00, 0.97–0.99, 0.95–0.97, and so on down to 0.85). A five-level scoring criterion was developed, and the LLM was prompted with the following instructions to rescore the selected samples. This allowed us to compare the consistency and differences between the two sets of scores.

\begin{tcolorbox}[colback=gray!20, colframe=gray!100, sharp corners, leftrule={3pt}, rightrule={0pt}, toprule={0pt}, bottomrule={0pt}, left={2pt}, right={2pt}, top={3pt}, bottom={3pt}]
{\footnotesize
Please evaluate the synonymy between the following two sentences on a scale from 5 (completely synonymous) to 1 (not synonymous) and provide the score along with a brief explanation:
\{sentence1\} and \{sentence2\}.

\textbf{Scoring Criteria as follows:}

\textbf{Completely Synonymous (5 points):} The core meaning of both sentences is identical, with only differences in expression.

\textbf{Highly Synonymous (4 points):} The core meaning is the same, but there are slight extensions, omissions, or differences in emphasis.

\textbf{Partially Synonymous (3 points):} The core meaning overlaps partially, but there are significant differences in focus or interpretation.

\textbf{Low Synonymy (2 points):} Only some keywords or parts of the content are similar, but the overall meaning is unrelated.

\textbf{Not Synonymous (1 point):} The core meanings of the two sentences are entirely different, with no semantic connection.
}
\end{tcolorbox}


\begin{figure}[t]
    \centering  
    \includegraphics[width=\textwidth, 
        trim={1cm 8.5cm 1cm 1cm},  
        clip     
    ]{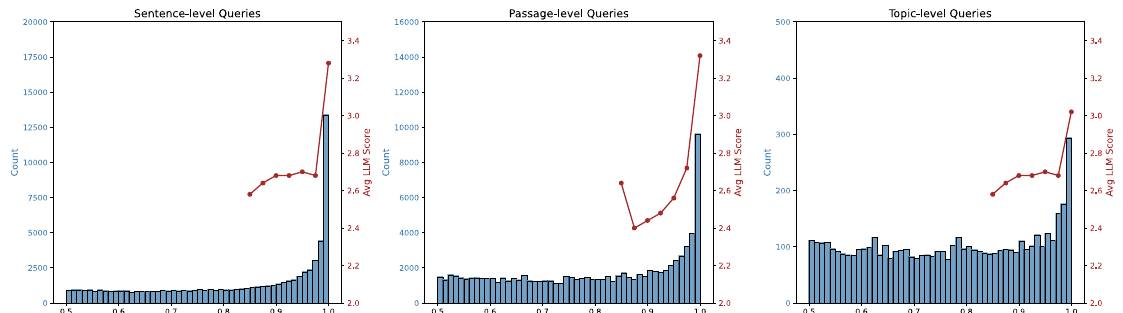}
    \caption{The distribution of similarity labels from the re-ranker and the average LLM score across different intervals.}
    \label{fig4}
\end{figure}
\begin{figure}[t!]
    \centering  
    \includegraphics[width=\textwidth, 
        trim={0cm 8.5cm 0cm 1cm},  
        clip     
    ]{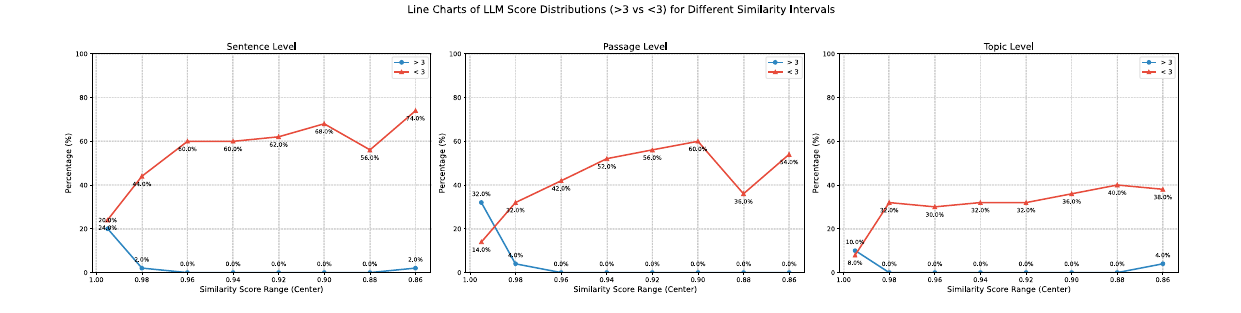}
    \caption{Line charts showing the distributions of LLM scores (>3 vs. <3) across different similarity intervals. We can probe the approximate percentage of false positive and false negative rate in the relevant labels from our indirect positives mining.
}
    \label{fig5}
\end{figure}
\subsubsection{Results and Analysis}

As shown in the figure \ref{fig4}, we observe that the similarity distribution from the reranker peaked in the 0.99 to 1 interval. This means that the 0.99 cut-off is an important feature of the model's discrimination ability.
At the same time, the average LLM scores for samples in this interval were significantly higher than others. Therefore, it is reasonable we set the threshold at 0.99. Our manual review to verify the scoring results of the LLM showed that over 90\% of the LLM's judgments were accepted by human evaluators. Considering potential human errors and the inherent limitations of the task, which may not be scored absolutely, we can assume LLM judgments as the ground truth, enabling us to estimate the approximate false positive and false negative rates in the relevant labels from our indirect positives mining. 


As shown in the figure \ref{fig5}, the false negative rate (i.e., the proportion of samples with a similarity score below 0.99 threshold but with a rating >3) remains low across all levels (4\%), demonstrating effective retention of high-quality relevance samples. 
For the false positive rate (i.e., the proportion of samples with a similarity score above threshold 0.99 but with a rating <3), the topic level (10\%) and passage level (14\%) performed well, while the sentence level was relatively higher (24\%). This stratified difference can be attributed to the characteristics of LLM in synonymy judgment. Through manual review, we find that LLMs apply stricter semantic detail comparisons than human evaluators, particularly at the sentence level. Since sentence-level queries often contain more specific details, minor semantic differences are amplified in LLM scoring, leading to a higher false positive rate. In contrast, the topic-level and passage-level queries, with their higher degree of abstraction and greater semantic tolerance, achieved a relatively lower false positive rate.

\subsection{Evaluation of FinCPRG}
Our evaluation includes experiments where FinCPRG is utilized both as a benchmark and a training dataset. The first component directly compares the evaluation results before and after replacing the benchmark in open-source financial retrieval evaluations with FinCPRG. The second component fine-tunes open-source models on our FinCPRG and evaluates their performance on other financial retrieval benchmarks to assess the pipeline's potential of generating training sets for low-resource domains.

\begin{table}[b]
\centering
\caption{
We evaluate models on FinCPRG tasks using Recall@10, maximally aligning with the metric (Recall@k) adopted in FIR-Bench for consistent benchmarking.
}
\label{tab2}
\begin{adjustbox}{max width=0.8\textwidth}
\begin{tabular}{lcccccc}
\toprule
\textbf{Model} & \multicolumn{6}{c}{\textbf{FinCPRG}} \\
\cmidrule(lr){2-7}
 & \textbf{All} & \textbf{Sentence} & \textbf{Sentence-mined} & \textbf{Passage} & \textbf{Passage-mined} & \textbf{Topic} \\
\midrule
\textbf{bge-base} & 0.719 & 0.696 & 0.649 & 0.865 & 0.695 & 0.381 \\
\textbf{bce-embedding-base} & 0.757 & 0.743 & 0.658 & 0.889 & 0.639 & 0.232 \\
\textbf{FinRetriever-base} & 0.780 & 0.768 & 0.715 & 0.896 & 0.704 & 0.391 \\
\textbf{FinRetriever-large} & 0.795 & 0.787 & 0.724 & 0.893 & 0.708 & 0.349 \\
\bottomrule
\end{tabular}
\end{adjustbox}
\end{table}

\begin{table}[t]
\centering
\caption{We evaluate models on FinCPRG tasks using NDCG@10, aligning with the metric adopted in FinMTEB for consistent benchmarking.}
\label{tab3}
\begin{adjustbox}{max width=0.8\textwidth}
\begin{tabular}{lcccccc}
\toprule
\textbf{Model} & \multicolumn{6}{c}{\textbf{FinCPRG}} \\
\cmidrule(lr){2-7}
 & \textbf{All} & \textbf{Sentence} & \textbf{Sentence-mined} & \textbf{Passage} & \textbf{Passage-mined} & \textbf{Topic} \\
\midrule
\textbf{bge-base} & 0.588 & 0.541 & 0.440 & 0.745 & 0.447 & 0.277 \\
\textbf{bge-large} & 0.546 & 0.506 & 0.415 & 0.678 & 0.402 & 0.279 \\
\textbf{bge-m3} & 0.706 & 0.667 & 0.527 & 0.832 & 0.475 & 0.265 \\
\textbf{all-MiniLM} & 0.116 & 0.114 & 0.087 & 0.120 & 0.052 & 0.022 \\
\textbf{multilingual-MiniLM} & 0.266 & 0.252 & 0.195 & 0.312 & 0.149 & 0.075 \\
\textbf{text2vec-base} & 0.403 & 0.392 & 0.304 & 0.430 & 0.237 & 0.129 \\
\bottomrule
\end{tabular}
\end{adjustbox}
\end{table}

\begin{figure}[h]
    \centering  
    \includegraphics[width=\textwidth, 
        trim={1cm 5.5cm 2cm 1cm},  
        clip     
    ]{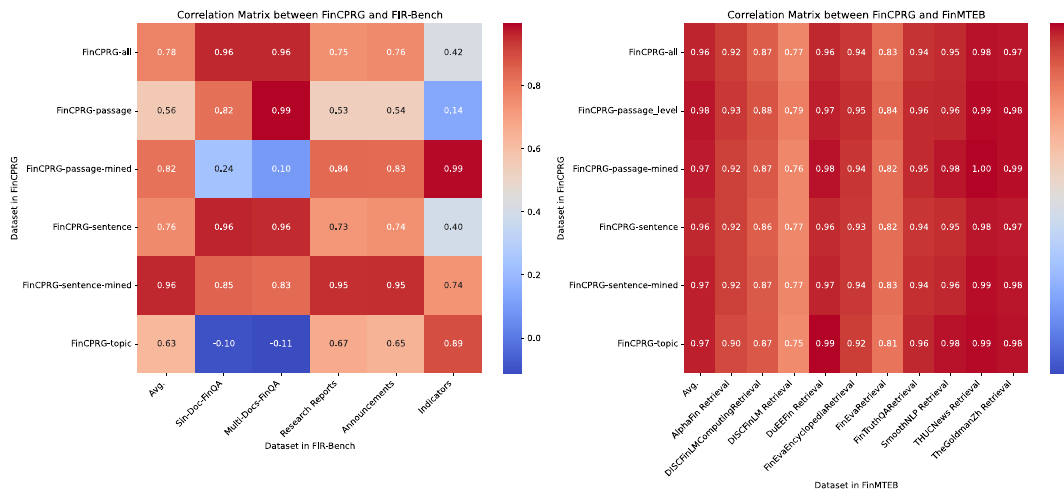}
    \caption{Correlation matrices of FinCPRG with two financial retrieval benchmarks, FIR-Bench (left) and FinMTEB (right). Each value of the correlation matrix represents the Pearson correlation between the test results of certain subset of our FinCPRG and the test results of certain subset of the real financial benchmark.}
    \label{fig6}
\end{figure}

\subsubsection{Serving as a Benchmark}
To ensure the fairness of comparations, we evaluated on FinCPRG while aligning with the original baselines and metric settings with two Chinese financial retrieval benchmarks, i.e FinMTEB\cite{tang2025finmtebfinancemassivetext} (BGE-Large-zh-v1.5, Paraphrase-multilingual-MiniLM-L12-v2, All-MiniLM-L12-v2, and BGE-M3) and FIR-Bench\cite{xu2025finbert2} (bge-base-zh-v1.5, bce-embedding-base-v1, FinRetriever-base, and FinRetriever-large). The evaluation was implemented based on Cocktail\cite{dai2024cocktail}, which includes a user-friendly evaluation tool. The results are shown in table \ref{tab2} and table \ref{tab3}.



Then we calculated the Pearson correlation coefficient for the evaluation results of shared models between FinCPRG and FIR-Bench as well as the results of shared models between FinCPRG and FinMTEB. As shown in the figure \ref{fig6}, we observe that most values are high, which reflects the utility consistency between FinCPRG and the real financial benchmark. Due to the narrow performance gap among the officially selected models in FIR-Bench and the inconsistent benchmarking metrics (i.e., varying $ k $ in Recall@k) used in FIR-Bench compared to our dataset, the correlation coefficients show fluctuations. In contrast, the large performance gap among the officially selected models in FinMTEB results in highly consistent test outcomes across different datasets.

\begin{table}[t]
\centering
\caption{Performance comparison of models fine-tuned on FinCPRG and evaluated on various datasets within the FinMTEB benchmark. The results are reported using NDCG@10. The suffix "ft" is the abbreviation for "fine-tuned".
}
\label{tab4}
\begin{adjustbox}{max width=\textwidth}
\begin{tabular}{llcccccccccc}
\toprule
\multirow{2}{*}{\textbf{Model}} & \multirow{2}{*}{\textbf{Avg.}} & \multicolumn{9}{c}{\textbf{Dataset in FinMTEB}} \\
\cmidrule{3-12}
 &  & \textbf{AlphaFin} & \textbf{DISCFinLLM-C} & \textbf{DISCFinLLM} & \textbf{DuEEFin} & \textbf{FinEva-E} & \textbf{FinEva} & \textbf{FinTruthQA} & \textbf{SmoothNLP} & \textbf{THUCNews} & \textbf{TheGoldmanZh} \\
\midrule
\textbf{bge-base} & 0.458 & 0.676 & 0.941 & 0.671 & 0.018 & 0.539 & 0.962 & 0.370 & 0.105 & 0.187 & 0.108 \\
\textbf{bge-base-ft} & 0.640 & 0.678 & 0.926 & 0.660 & 0.223 & 0.860 & 0.965 & 0.419 & 0.632 & 0.604 & 0.429 \\
\textbf{bge-large} & 0.607 & 0.664 & 0.948 & 0.672 & 0.243 & 0.432 & 0.955 & 0.433 & 0.695 & 0.588 & 0.442 \\
\textbf{bge-large-ft} & 0.619 & 0.694 & 0.915 & 0.687 & 0.222 & 0.785 & 0.962 & 0.404 & 0.619 & 0.542 & 0.365 \\
\textbf{MiniLM} & 0.379 & 0.487 & 0.807 & 0.646 & 0.036 & 0.316 & 0.878 & 0.242 & 0.049 & 0.140 & 0.195 \\
\textbf{MiniLM-ft} & 0.533 & 0.628 & 0.710 & 0.879 & 0.093 & 0.752 & 0.954 & 0.311 & 0.333 & 0.403 & 0.269 \\
\bottomrule
\end{tabular}
\end{adjustbox}
\end{table}

\subsubsection{Serving as Training Dataset}
To evaluate the effectiveness of synthetic datasets as a training dataset, we conducted fine-tuning experiments using FinCPRG and tested the results on FinMTEB. Specifically, we fine-tuned three models: bge-base-zh-v1.5, bge-large-zh-v1.5, and paraphrase-multilingual-MiniLM-L12-v2. The fine-tuning process employed the CoSENT loss function and mean pooling strategy without hard negative mining. The experimental results are summarized in the table \ref{tab4}.

Overall, all three models achieved an average performance improvement across ten evaluation datasets after fine-tuning. The BGE model’s strong performance on certain financial retrieval evaluation sets can be attributed to its good generalization capabilities or possibly to data leakage in some of the evaluation datasets. We observed that if the model initially performed well on a particular dataset, its performance would show slight fluctuations after our fine-tuning. Conversely, if the model has not been adequately trained, fine-tuning would lead to a significant improvement in its performance. Notably, the models that initially exhibited lower performance experienced the most significant gains. For instance, MiniLM achieved an average improvement of approximately 15\%. This indicates that our pipeline is well suited for low-resource areas to improve domain capabilities.

\section{Discussions}
\subsection{Estimation about computational cost and scalability}

We evaluate the computational cost of each stage in our multi-stage pipeline, where costs comprise inference from three model types: BERT-based models, LLMs, and reranker models.  For the raw corpus, we assume $n$ documents with an average of $t$ tokens each, resulting in $NC_1$ sentence-level chunks and $NC_2$ passage-level chunks total, where $NC_1$ is calculated as \(\frac{nt}{100}\) and $NC_2$ as \(\frac{nt}{500}\).

BERT inference costs for document cleaning ($\frac{t}{500} \times n$), industry labeling ($n$), and topic clustering ($n$) yield a total cost proportional to $NC_2$.

LLM inference costs for query generation scale with token count: intra-document generation ($2 \times t \times n$) and cross-document generation ($n \times L_2 \times 3$), making the total cost approximately proportional to corpus tokens.

Reranker mining costs vary in different parts. Sentence-level: $n \times (\frac{t}{100})^2 \propto NC_1$
Passage-level: Within topic subtrees with $m$ clusters, cost scales as $\frac{n^2 t^2}{m 500^2} \propto NC_2^{1.5}$ (using empirical formula for clustering $m = \sqrt{n}$).
Topic-level: $m \times \frac{n}{m} \times \frac{t}{500} \propto NC_2$.

Overall, the pipeline demonstrates favorable scalability, with costs scaling linearly to sub-quadratically with corpus size.

\subsection{Limitations and Future Work}
\subsubsection{Coverage and Scale of Raw Data}
Our dataset comprises 39w financial research reports spanning 2022 to January 2024, distributed across 20+ report categories. Despite this broad coverage, the relatively limited samples per category (<100) may not fully capture domain-specific nuances. Future research could benefit from focused sampling within specific subdomains of finance and increasing sample quantities in high-priority categories.

\subsubsection{Stability of Pipeline}
The multi-stage nature of our approach—incorporating various language models including API of commercial LLMs, complex manual sequential operations, and vast hyperparameter configurations—introduces inherent variability. This complexity affects model output consistency and cross-run reproducibility. Addressing these stability issues requires systematic evaluation of model robustness and development of standardized benchmarking procedures.

\subsubsection{Quality Constraints in Pipeline}
The quality of generated data faces several inherent inadequacies in the design of the pipeline, such as limited context in cross-doc query synthesis, insufficient utilization of doc metadata and inadequate incorporation of temporal context. Additionally, there can be non-conforming outputs at every step of the pipeline, such as wrong reasoning of BERT-based cleaning and labeling, inherited biases from LLM generation, bad cases in rule-based entity completion, and threshold-based relevance judgment. Future improvements could focus on enhanced prompt engineering, model optimization, and development of automated verification mechanisms for synthetic data.

\section{Conclusion}

We proposed a bidirectional generation pipeline to construct hierarchical queries and enrich relevance labels for financial passage retrieval. Using this method, we created FinCPRG, a financial Chinese passage retrieval generated dataset with hierarchical queries and rich relevance annotations. Evaluations of mined relevance labels and experiments as both a benchmark and training dataset confirmed our proposed pipeline's effectiveness, showcasing its potential for generating effective passage retrieval datasets for low-resource domains.

\section{Acknowledgement}

This work was supported in part by the National Key Research and Development Program of China under Grant 2022YFC3303301, Grant 2023YFC3305402, Grant 2023YFC3305401 and in part by the National Natural Science Foundation of China (Nos.62302059 and 62172053).

%
%
%
\bibliographystyle{splncs04}

%





\end{document}